\documentclass{amsart}

\usepackage{graphicx}
\usepackage{amssymb}
\usepackage[font={footnotesize,it}]{caption}
\usepackage{url}
\usepackage{mathtools}
\usepackage{braket}
\usepackage{qcircuit}
\usepackage{hyperref}

\newtheorem{theorem}{Theorem}
\newtheorem{proposition}[theorem]{Proposition}
\newtheorem{lemma}[theorem]{Lemma}
\newtheorem{corollary}[theorem]{Corollary}
\theoremstyle{definition}

\newtheorem{assumption}[theorem]{Assumption}
\theoremstyle{remark}
\newtheorem{remark}[theorem]{Remark}
\numberwithin{theorem}{section}
\numberwithin{equation}{section}

\DeclareMathOperator{\polylog}{polylog}
\DeclareMathOperator{\Rop}{\mathcal{R}}
\DeclarePairedDelimiter\ceil{\lceil}{\rceil}
\DeclarePairedDelimiter\floor{\lfloor}{\rfloor}
\newcommand{\be}{\begin{equation}}
\newcommand{\ee}{\end{equation}}
\newcommand{\ben}{\begin{equation*}}
\newcommand{\een}{\end{equation*}}
\newcommand{\err}{\varepsilon}

\begin{document}

\title[Exponential sums and $\zeta(s)$ on a quantum computer]{Quantum algorithms for exponential sums and the evaluation of the Riemann zeta function}
\author{Sandeep Tyagi}
\email{tyagi\_sandeep@yahoo.com}
\address{London, United Kingdom}
\date{\today}

\begin{abstract}
We give quantum algorithms for estimating weighted exponential sums of the form
\begin{equation*}
S(f,w,N)= \sum_{k=0}^{N-1} w_k\, e^{2 \pi i f(k)},
\qquad w_k \ge 0,\quad \sum_{k=0}^{N-1} w_k =1,
\end{equation*}
where $N=2^{n}$ may be exponentially large. Under two explicitly stated oracle assumptions---(i) the prefix sums $\sum_{k<M} w_k$ admit efficient reversible circuits, so that the state $\sum_k \sqrt{w_k}\,\ket{k}$ can be prepared by the Grover--Rudolph method, and (ii) $f$ admits an efficient fixed-point implementation---we show that $S(f,w,N)$ can be estimated to additive error $\err$, with failure probability at most $\gamma$, at a cost of $O(\err^{-1}\log(1/\gamma))$ uses of the two oracles and $\polylog(N)$ gates per use. The algorithm applies amplitude estimation together with an explicit accounting of state-preparation and oracle costs---so all bounds quoted are full gate complexities, not merely query complexities---and achieves a quadratic saving in $1/\err$ over classical Monte Carlo sampling of the same sum.

As the main application we analyze the evaluation of the Riemann zeta function via the Riemann--Siegel formula. For $s=\sigma+it$ in the critical strip we prove that $\zeta(s)$ can be estimated to additive accuracy $\delta$ with
\begin{equation*}
O\!\left( \frac{t^{(1-\sigma)/2}}{\delta}\, \polylog(t/\delta)\log(1/\gamma) \right)
\end{equation*}
quantum gates; on the critical line $\sigma = 1/2$ the cost is $O(t^{1/4}\delta^{-1}\polylog(t/\delta))$. For fixed accuracy this improves on the $t^{1/2}$ scaling of the Riemann--Siegel formula and on the best rigorous classical algorithm, which runs in time $t^{4/13+o(1)}$. We delineate precisely the accuracy regime in which the quantum algorithm is advantageous, show that Hiary-type block decompositions of the main sum cannot improve the quantum query complexity, and explain why a $\polylog(t)$ quantum algorithm does not follow from these techniques: amplitude estimation yields additive precision on a normalized amplitude, and undoing the normalization costs a factor equal to the $\ell^{1}$ mass $\sum_{k\le N} k^{-\sigma} = \Theta(t^{(1-\sigma)/2})$ of the main sum. We also give a method to estimate the magnitude $\lvert \sum_k a_k \rvert$ of the amplitude sum of an arbitrary efficiently preparable $n$-qubit state, and review the amplitude- and phase-estimation subroutines on which all of the above rely.
\end{abstract}

\maketitle

\section{Introduction}
\label{sec:intro}

An exponential sum (ES) with weights $w = (w_0,\dots,w_{N-1})$ and phase function $f$ is
\begin{align}
S(f, w, N) = \sum_{k=0}^{N-1} w_k \exp( 2 \pi i f(k) ),
\qquad w_k \ge 0, \quad \sum_{k=0}^{N-1} w_k = 1,
\label{eq:expsum}
\end{align}
where $f:\{0,\dots,N-1\}\to\mathbb{R}$ is a real function. Sums of this shape---most commonly with uniform weights and a polynomial phase $f = p_d$ of degree $d$---are ubiquitous in analytic number theory. They control bounds for the Riemann zeta (RZ) function $\zeta(s)$ in the critical strip, appear in the theory of Diophantine equations and congruences, and underlie the fastest known methods for evaluating $\zeta(1/2+it)$ at large height $t$ \cite{titchmarsh1986theory, hiary2011fast}. They have also been proposed as a route to integer factorization, in which the factors of an integer appear as peaks of a Gauss-type sum with quadratic phase \cite{woelk2011factorization, merkel2011factorization, shor1999polynomial}.

Evaluated term by term, \eqref{eq:expsum} costs $\Theta(N)$ operations. For special phases much better is possible classically: quadratic Gauss sums admit closed forms, and Hiary \cite{hiary2011nearly} showed that the truncated theta function (quadratic phase, uniform weights) can be computed to within $\pm N^{-\lambda}$ in $\polylog(N)$ time. For cubic and higher-degree phases no comparably fast classical method is known, and this is the principal obstruction to faster rigorous evaluation of $\zeta(s)$ at large height \cite{hiary2011fast}.

\subsection{Contributions}
This paper analyzes what a quantum computer (QC) can and cannot do for such sums, with explicit oracle assumptions and error accounting. Our results are as follows.

\begin{enumerate}
\item \emph{Weighted exponential sums} (Theorem~\ref{thm:es}). Given (i) a state-preparation oracle for $\ket{\psi_w}=\sum_k \sqrt{w_k}\ket{k}$, realizable by the Grover--Rudolph construction \cite{grover2002creating} whenever the prefix sums $S_w(M)=\sum_{k<M}w_k$ are efficiently computable, and (ii) a fixed-point phase oracle for $f$, the sum \eqref{eq:expsum} can be estimated to additive error $\err$ with $O(\err^{-1}\log(1/\gamma))$ oracle calls. The method is amplitude estimation \cite{brassard2002quantum} applied to a controlled rotation of an ancilla qubit; it should be viewed as a quantum mean-estimation algorithm in the sense of \cite{montanaro2015quantum, abrams1999fast}, specialized to number-theoretic sums. The advantage over classical sampling of the same quantity is quadratic in $1/\err$ (Remark~\ref{rem:montecarlo}).

\item \emph{Riemann zeta function} (Theorem~\ref{thm:zeta}). Combining Theorem~\ref{thm:es} with the Riemann--Siegel (RS) formula, $\zeta(\sigma+it)$ can be estimated to additive accuracy $\delta$, for any fixed $\sigma$ in the critical strip, with $O(t^{(1-\sigma)/2}\delta^{-1}\polylog(t/\delta)\log(1/\gamma))$ gates; in particular $O(t^{1/4}\delta^{-1}\polylog(t/\delta))$ on the critical line. Table~\ref{tab:compare} compares this with the classical state of the art. We emphasize what the theorem does \emph{not} say: the dependence on the accuracy is $\delta^{-1}$, not $\polylog(1/\delta)$, so the algorithm does not compute $\zeta$ to within $t^{-\lambda}$ in time $t^{o(1)}$. Remark~\ref{rem:crossover} quantifies the exact crossover: the quantum algorithm improves on the best classical $t^{4/13+o(1)}$ method precisely when the demanded accuracy satisfies $\delta \gg t^{-3/52}$, which comfortably covers the accuracy needed to separate and count zeros on the critical line (Remark~\ref{rem:zeros}).

\item \emph{No further gain from block decomposition} (Proposition~\ref{prop:blocks}). The classical $t^{1/3}$ and $t^{4/13}$ algorithms of \cite{hiary2011fast} split the RS main sum into blocks that are then evaluated by fast theta-function algorithms. We show that, on a quantum computer, any such decomposition can only increase the worst-case query complexity: amplitude estimation applied directly to the full sum already realizes the square-root savings that blocking is designed to capture.

\item \emph{Magnitude of an amplitude sum} (Proposition~\ref{prop:mag}). For any state $\ket{\psi}=\sum_k a_k \ket{k}$ prepared by a known circuit, the quantity $N^{-1/2}\lvert\sum_k a_k\rvert$ can be estimated to additive error $\err$ with $O(\err^{-1}\log(1/\gamma))$ uses of the circuit. Applied to $a_k=\sqrt{w_k}e^{2\pi i f(k)}$ this estimates the \emph{amplitude-weighted} sum $\sum_k \sqrt{w_k}e^{2\pi i f(k)}$, but only in magnitude and with an unavoidable $\sqrt{N}$ normalization penalty, which we make explicit.

\item A self-contained, corrected review of the amplitude- and phase-estimation subroutines used above (Section~\ref{sec:phase}), including QFT-based estimation and QFT-free estimation with classical post-processing \cite{kitaev1995quantum, svore2013faster, wiebe2016efficient, suzuki2020amplitude, aaronson2020quantum}.
\end{enumerate}

\begin{table}
\centering
\footnotesize
\begin{tabular}{lll}
\hline
Method & Cost to accuracy $\delta$ at $\sigma=\tfrac12$ & Model \\
\hline
Euler--Maclaurin \cite{rubinstein2005computational} & $t^{1+o(1)}\polylog(1/\delta)$ & classical, rigorous \\
Riemann--Siegel \cite{edwards1974riemannaes, gabcke1979neue} & $t^{1/2+o(1)}\polylog(1/\delta)$ & classical, rigorous \\
Hiary \cite{hiary2011fast} & $t^{4/13+o(1)}\,\mathrm{poly}(\lambda)$, $\delta = t^{-\lambda}$ & classical, rigorous \\
This work (Thm.~\ref{thm:zeta}) & $t^{1/4}\,\delta^{-1}\polylog(t/\delta)$ & quantum, oracle costs included \\
\hline
\end{tabular}
\medskip
\caption{Cost of evaluating $\zeta(1/2+it)$ to additive accuracy $\delta$. The quantum cost is favorable for fixed or polynomially small $\delta \gg t^{-3/52}$; see Remark~\ref{rem:crossover}.}
\label{tab:compare}
\end{table}

\subsection{Background on \texorpdfstring{$\zeta(s)$}{zeta}}
Throughout, $s=\sigma+it$ with $\sigma, t \in \mathbb{R}$, and the critical strip is $0<\sigma<1$. For $\Re(s)>1$ the zeta function is defined by the absolutely convergent series
\begin{align}
\zeta(s) = \sum_{k=1}^{\infty} k^{-s},
\end{align}
and extends to a meromorphic function on $\mathbb{C}$ with a single simple pole at $s=1$ \cite{edwards1974riemannaes, titchmarsh1986theory}. The distribution of prime numbers is governed by the non-trivial zeros of $\zeta$, all of which lie in the critical strip. The Riemann hypothesis (RH)---that every non-trivial zero has $\sigma=1/2$---has resisted proof for more than a century and a half, and a large body of results in number theory is conditional on it \cite{montgomery2017exploring}.

One line of attack, going back to Riemann's own hand computations \cite{edwards1974riemannaes}, is numerical: a single zero off the critical line would disprove RH, and large-scale verifications also supply statistical information about the zeros. RH has been verified for the first $10^{13}$ zeros \cite{gourdon20041013} and, rigorously, for all zeros with $0< t \le 3\cdot 10^{12}$ \cite{platt2021riemann}; values of $\zeta(1/2+it)$ have been computed at heights near $t=10^{36}$ \cite{bober2018new}. All such computations at large height rest on the RS formula, whose main sum has $N=\floor{\sqrt{t/2\pi}}$ terms, so that direct evaluation costs $\Theta(t^{1/2})$. Hiary \cite{hiary2011fast} showed how to reduce this: partition the main sum into consecutive blocks, approximate each block by an exponential sum with polynomial phase of low degree, and evaluate those by fast theta-function algorithms \cite{hiary2011nearly}. Quadratic phases give overall cost $t^{1/3+o(1)}$ and cubic phases give $t^{4/13+o(1)}$; the absence of fast algorithms for degree $\ge 4$ blocks halts further progress along these lines. Alternative formulas with more tractable error terms are developed in \cite{hiary2016alternative, arias2011high}.

\subsection{Related work}
Van Dam and Seroussi \cite{van2002efficient} gave efficient quantum algorithms for estimating Gauss sums over finite fields; apart from this, quantum algorithms for number-theoretic exponential sums appear largely unexplored. The estimation technique we use belongs to the family of amplitude-estimation-based mean and integral estimators \cite{brassard2002quantum, abrams1999fast, montanaro2015quantum}, with state preparation via Grover--Rudolph \cite{grover2002creating} and function-controlled rotations of the kind used in quantum risk analysis and option pricing \cite{woerner2019quantum, stamatopoulos2019option}. Quantum circuits for the fixed-point evaluation of $\ln$ and other transcendental functions are given in \cite{wang2020quantum, haner2018optimizing}.

\subsection{Organization}
Section~\ref{sec:prelim} fixes the computational model, the oracle assumptions, and the amplitude-estimation toolbox. Section~\ref{sec:gen} proves the exponential-sum theorem. Section~\ref{sec:rot} describes circuits for function-controlled rotations and their errors. Section~\ref{sec:riemann} collects the classical Euler--Maclaurin and RS formulas with the remainder bounds we need. Section~\ref{sec:qc} proves the zeta theorem and the no-gain result for block decompositions. Section~\ref{sec:mag} treats the magnitude of amplitude sums. Section~\ref{sec:phase} reviews phase and amplitude estimation. Section~\ref{sec:conclusion} concludes.

\section{Preliminaries}
\label{sec:prelim}

\subsection{Model and notation}
We work in the quantum circuit model over a fixed universal gate set; ``gate complexity'' counts elementary gates, and single-qubit rotations by arbitrary angles are counted as single gates (compiling them to a discrete fault-tolerant gate set costs a further multiplicative $O(\log(1/\err))$ by Solovay--Kitaev \cite{kitaev2002classical, nielsen2002quantum}, which is absorbed into $\polylog$ factors). For $N = 2^n$, $\ket{k}_n$ denotes the $n$-qubit computational basis state encoding $k \in \{0, \dots, N-1\}$. Real quantities are represented in fixed point: a register $\ket{q_{m_1}, \dots, q_0 . q_{-1}, \dots, q_{-m_2}}$ encodes $\sum_{i=-m_2}^{m_1} q_i 2^{i}$ with $q_i \in \{0,1\}$. We write
\be
\Rop(\alpha) := e^{-i\pi \alpha Y} =
\begin{pmatrix} \cos \pi\alpha & -\sin \pi\alpha \\ \sin \pi\alpha & \cos \pi\alpha \end{pmatrix},
\qquad \Rop(\alpha)\ket{0} = \cos(\pi\alpha)\ket{0} + \sin(\pi\alpha)\ket{1},
\label{eq:rgate}
\ee
so that $\Rop(\alpha)\Rop(\beta) = \Rop(\alpha+\beta)$.

\subsection{Oracle assumptions}

\begin{assumption}[Weight oracle]
\label{ass:weights}
The weights $w_k \ge 0$, $\sum_{k=0}^{N-1} w_k = 1$, are such that the prefix sums $S_w(M) = \sum_{k=0}^{M-1} w_k$ (equivalently, the conditional interval probabilities arising in binary bisection) are computable to $b$ bits of precision by classical circuits of size $T_w(b, n) = \mathrm{poly}(b, n)$.
\end{assumption}

\begin{assumption}[Phase oracle]
\label{ass:phase}
The phase $f$ is computable in fixed point: there is a reversible circuit of size $T_f(b,n)=\mathrm{poly}(b,n)$ mapping $\ket{k}\ket{0} \mapsto \ket{k}\ket{\tilde f(k)}$, where $\tilde f(k)$ is a $b$-bit fixed-point number with $|\tilde f(k) - f(k) \bmod 2| \le 2^{-b}$. (Only $f \bmod 2$ is relevant, since the rotation \eqref{eq:rgate} has period $2$ in $\alpha$.)
\end{assumption}

Both assumptions hold for the applications in this paper: for the zeta weights $w_k \propto k^{-\sigma}$ the prefix sums are computable to precision $2^{-b}$ in $\mathrm{poly}(b, \log N)$ time by the Euler--Maclaurin formula (Section~\ref{sec:riemann}), and phases of the form $f(k) = -\tfrac{t}{2\pi}\ln k$ or $f = p_d(k)$ are computable by standard fixed-point arithmetic \cite{haner2018optimizing, wang2020quantum}.

\begin{lemma}[State preparation; Grover--Rudolph \cite{grover2002creating}]
\label{lem:gr}
Under Assumption~\ref{ass:weights}, for every $\eta \in (0,1)$ there is a quantum circuit $W$ of size $O\bigl(n \,(T_w(b,n) + \mathrm{poly}(b))\bigr)$ with $b = O(\log (n/\eta))$ such that
\ben
\Bigl\| W\ket{0}_n - \sum_{k=0}^{N-1} \sqrt{w_k}\, \ket{k}_n \Bigr\| \le \eta .
\een
\end{lemma}

\begin{proof}[Proof sketch]
Proceed by binary bisection: after $j$ levels the register holds $\sum_{I} \sqrt{w(I)}\ket{I}$ over the $2^j$ dyadic intervals $I$ of length $2^{n-j}$, with $w(I)$ the total weight of $I$. Refining one level applies, controlled on $I$, the rotation $\Rop(\theta_I/\pi)$ with $\theta_I = \arccos\sqrt{w(I_{\mathrm{left}})/w(I)}$ to a fresh qubit. The angle is computed reversibly into an ancilla register---two prefix-sum evaluations, a subtraction, a division, a square root and an arccosine, all in $b$-bit fixed point and executed \emph{coherently} over the superposition of interval labels $I$---then consumed bit by bit as in Figure~\ref{fig:rotCircuit} and uncomputed. Note that the global normalization of $w$ cancels from the conditional ratios $w(I_{\mathrm{left}})/w(I)$, so it is never needed inside the quantum circuit; it enters only the classical post-processing (Corollary~\ref{cor:unnormalized}). Each of the $n$ levels introduces amplitude error $O(2^{-b})$, and the total $\ell^2$ error is $O(n 2^{-b} )$; choose $b$ accordingly. (Intervals of zero or negligible conditional weight are handled by fixing the rotation angle to $0$ or $\pi/2$.)
\end{proof}

\subsection{Amplitude estimation}
We use the standard amplitude-estimation theorem. Let $\mathcal{A}$ be a unitary on $n+1$ qubits with
\be
\mathcal{A}\ket{0}_{n+1} = \sin(\theta_p)\ket{\Psi_1} + \cos(\theta_p)\ket{\Psi_0},
\label{eq:ampl}
\ee
where $\ket{\Psi_1}$ (the ``good'' component) lies in the subspace flagged by a designated qubit and $p = \sin^2\theta_p$, $\theta_p \in [0, \pi/2]$, is the probability of measuring the flag.

\begin{theorem}[Brassard--H{\o}yer--Mosca--Tapp {\cite[Thm.~12]{brassard2002quantum}}]
\label{thm:ae}
For every integer $M \ge 1$ there is a quantum algorithm using $M$ applications of the Grover iterate $Q = -\mathcal{A} S_0 \mathcal{A}^{-1} S_{\chi}$ (see Section~\ref{sec:phase}) that outputs $\tilde p \in [0,1]$ satisfying
\ben
| \tilde p - p | \le \frac{2\pi \sqrt{p(1-p)}}{M} + \frac{\pi^2}{M^2}
\een
with probability at least $8/\pi^2$.
\end{theorem}

\begin{lemma}[Median amplification]
\label{lem:median}
Repeating the estimator of Theorem~\ref{thm:ae} independently $r = O(\log(1/\gamma))$ times and taking the median yields the same accuracy with failure probability at most $\gamma$.
\end{lemma}

\begin{proof}
Each run succeeds with probability $8/\pi^2 > 0.81 > 1/2$; the median fails only if at least half the runs fail, which by a Chernoff bound has probability $e^{-\Omega(r)}$.
\end{proof}

\section{Estimating weighted exponential sums}
\label{sec:gen}

\begin{theorem}
\label{thm:es}
Let $N = 2^n$, let $w$ satisfy Assumption~\ref{ass:weights} and $f$ satisfy Assumption~\ref{ass:phase}. For all $\err, \gamma \in (0,1)$ there is a quantum algorithm that outputs $\widehat S \in \mathbb{C}$ with
\ben
\bigl|\widehat S - S(f,w,N)\bigr| \le \err
\quad\text{with probability at least } 1 - \gamma,
\een
using $O(\err^{-1} \log (1/\gamma))$ applications of $W$, $W^{-1}$ and of the phase oracle, and
\ben
O\Bigl( \err^{-1}\, \log(1/\gamma) \bigl( T_w(b,n) + T_f(b,n) + \mathrm{poly}(b,n) \bigr) \Bigr),
\qquad b = O(\log (n/\err)),
\een
elementary gates in total. Under the stated polynomial bounds on $T_w, T_f$ this is $O(\err^{-1}\polylog(N/\err)\log(1/\gamma))$.
\end{theorem}

\begin{proof}
We estimate $\Re S$ and $\Im S$ separately, each to additive error $\err/\sqrt{2}$; the claim follows by combining them.

\emph{Circuit.} Let $c \in \{0, -\tfrac14\}$ be a constant offset. Define $\mathcal{A}_c = U_{f+c}\,(W \otimes I)$ acting on $\ket{0}_n\ket{0}$, where $U_{g}$ denotes the function-controlled rotation
\begin{align}
U_{g}\, \ket{k}_n \ket{0} = \ket{k}_n \bigl(\cos( \pi g(k)) \ket{0} + \sin(\pi g(k)) \ket{1} \bigr),
\label{eq:uf}
\end{align}
implemented from the phase oracle as described in Section~\ref{sec:rot} (compute $\tilde f(k)$ into a $b$-bit register, apply the controlled rotations of Figure~\ref{fig:rotCircuit}, uncompute). Then
\begin{align}
\mathcal{A}_c \ket{0}_{n}\ket{0} = \sum_{k=0}^{N-1} \sqrt{w_k}\, \ket{k}_n \bigl(\cos(\pi \tilde g(k)) \ket{0} + \sin(\pi \tilde g(k)) \ket{1} \bigr),
\qquad \tilde g = \tilde f + c .
\end{align}
Designating the ancilla value $\ket{0}$ as the good outcome, the good probability is
\begin{align}
p_c = \sum_{k=0}^{N-1} w_k \cos^2\bigl(\pi (\tilde f(k)+c)\bigr)
     = \frac12 + \frac12 \sum_{k=0}^{N-1} w_k \cos\bigl(2\pi (\tilde f(k)+c)\bigr).
\label{eq:pc}
\end{align}
Since $\cos(2\pi f) = \Re e^{2\pi i f}$ and $\cos(2\pi f - \pi/2) = \Im e^{2\pi i f}$,
\be
2 p_0 - 1 = \Re S(\tilde f, w, N), \qquad 2 p_{-1/4} - 1 = \Im S(\tilde f, w, N).
\label{eq:recover}
\ee

\emph{Truncation error.} By Assumption~\ref{ass:phase}, $|\tilde f(k) - f(k)| \le 2^{-b} \pmod 2$, and since $x \mapsto \cos^2(\pi x)$ is $\pi$-Lipschitz, \eqref{eq:pc} differs from its exact-$f$ value by at most $\pi 2^{-b}$; hence by \eqref{eq:recover} each of $\Re S, \Im S$ acquires a bias at most $2\pi\, 2^{-b} \le \err / 8$ for $b \ge \log_2(16\pi/\err)$.

\emph{Estimation error.} Apply Theorem~\ref{thm:ae} to $\mathcal{A}_c$ with $M = \ceil{16/\err}$: using $\sqrt{p(1-p)} \le 1/2$,
\ben
|\tilde p_c - p_c| \le \frac{\pi}{M} + \frac{\pi^2}{M^2} \le \frac{\pi \err}{16} + \frac{\pi^2 \err^2}{256} \le \frac{\err}{4},
\een
so $|(2\tilde p_c -1) - (2 p_c - 1)| \le \err/2$, and adding the truncation bias, each component of $S$ is estimated to within $\err/2 + \err/8 < \err/\sqrt 2$. By Lemma~\ref{lem:median}, $O(\log(1/\gamma))$ repetitions per component give total failure probability $\le \gamma$. Each Grover iterate uses $\mathcal{A}_c$, $\mathcal{A}_c^{-1}$ and two reflections, i.e., $O(1)$ calls to $W^{\pm 1}$ and the phase oracle plus $O(\mathrm{poly}(b,n))$ gates, which gives the stated complexity. State-preparation error from Lemma~\ref{lem:gr} is controlled by taking $\eta = O(\err)$ there and absorbing it in the constants, since an $\eta$-perturbation of $\mathcal{A}_c$ perturbs $p_c$ by $O(\eta)$.
\end{proof}

\begin{remark}[Anatomy of the cost]
\label{rem:anatomy}
It is worth displaying the structure of the bound in Theorem~\ref{thm:es} explicitly. Writing $T_W$ for the gate cost of the state-preparation circuit of Lemma~\ref{lem:gr}, $T_f$ for one phase-oracle call, and $T_R = O(n + b)$ for the reflections and controlled rotations of one Grover iterate,
\ben
T_{\mathrm{total}} \;=\; \underbrace{Q \cdot \bigl( T_W + T_f + T_R \bigr)}_{\text{quantum}} \;+\; T_{\mathrm{cl}},
\qquad Q = O\bigl( \err^{-1}\log(1/\gamma) \bigr),
\een
where $T_{\mathrm{cl}}$ is the classical post-processing (median statistics, and the normalization constant in Corollary~\ref{cor:unnormalized} below). Two points deserve emphasis. First, $W$ and $W^{-1}$ appear \emph{inside every} Grover iterate $Q = -\mathcal{A}S_0\mathcal{A}^{-1}S_\chi$, so the state-preparation cost multiplies $Q$; it is not a one-time cost, and a $\polylog$ total is meaningful only because Lemma~\ref{lem:gr} bounds $T_W$ under Assumption~\ref{ass:weights}. Second, no trigonometric function is ever evaluated numerically inside the circuit: the rotation by $\pi \tilde f(k)$ is applied bitwise with hard-wired angles (Figure~\ref{fig:rotCircuit}), so $T_f$ is purely the reversible arithmetic of Assumption~\ref{ass:phase}.
\end{remark}

\begin{corollary}[Unnormalized sums]
\label{cor:unnormalized}
Let $c_k \ge 0$ with $C = \sum_{k=0}^{N-1} c_k$ computable classically to relative error $O(\delta / C)$ in time $\polylog(N/\delta)$, and suppose $w_k = c_k / C$ satisfies Assumption~\ref{ass:weights}. Then $\sum_k c_k e^{2\pi i f(k)}$ can be estimated to additive error $\delta$ with probability $1-\gamma$ at cost
$O\bigl( (C/\delta) \polylog(N/\delta) \log(1/\gamma) \bigr)$.
\end{corollary}

\begin{proof}
Apply Theorem~\ref{thm:es} with $\err = \delta/(2C)$ and output $\widehat S \cdot \widehat C$ for a classical estimate $\widehat C$ of $C$.
\end{proof}

\begin{remark}[The $\ell^1$ mass is the price of denormalization]
\label{rem:l1}
Corollary~\ref{cor:unnormalized} is the source of all polynomial-in-$t$ factors later in the paper. Amplitude estimation delivers \emph{additive} precision on the normalized quantity $S(f,w,N) \in \overline{\mathbb{D}}$; recovering an unnormalized sum multiplies the error by $C = \|c\|_1$. No improvement is possible in general: estimating $|S|$ for uniform weights to additive $2^{-n/2}$ would distinguish states with inner products differing by that amount and requires $\Omega(2^{n/2})$ oracle calls by the optimality of Grover-type search \cite{brassard2002quantum}.
\end{remark}

\begin{remark}[Comparison with classical sampling]
\label{rem:montecarlo}
The normalized sum $S(f,w,N)$ is also estimable \emph{classically} in $\polylog(N)/\err^2$ time whenever one can sample $k \sim w$ (possible under Assumption~\ref{ass:weights} by inverse-transform sampling) and evaluate $f$: the empirical mean of $e^{2\pi i f(k)}$ over $O(\err^{-2}\log(1/\gamma))$ samples suffices, by Hoeffding's inequality. The quantum advantage of Theorem~\ref{thm:es} is therefore precisely the quadratic improvement $\err^{-2} \to \err^{-1}$ characteristic of amplitude-estimation-based mean estimation \cite{montanaro2015quantum}. This quadratic saving is what drives the improvement over classical algorithms for $\zeta$ in Section~\ref{sec:qc}, and Remark~\ref{rem:l1} is what prevents anything stronger.
\end{remark}

\begin{corollary}[Uniform weights: Weyl-type sums]
\label{cor:weyl}
Let $w_k = 1/N$. Then Assumption~\ref{ass:weights} is trivial and $W = H^{\otimes n}$, so $T_W = O(n)$ and no Grover--Rudolph machinery is needed. For any phase satisfying Assumption~\ref{ass:phase}---in particular any polynomial $p_d$ with fixed-point coefficients, implemented as in Section~\ref{sec:rot}---Theorem~\ref{thm:es} estimates $N^{-1}\sum_{k<N} e^{2\pi i p_d(k)}$ to additive error $\err$ at cost $O(\err^{-1} \polylog(N/\err) \log(1/\gamma))$. Equivalently, the Weyl sum $\sum_{k<N} e^{2\pi i p_d(k)}$ is obtained to additive error $\delta$ at cost $O((N/\delta)\polylog(N/\delta))$; when the sum exhibits square-root cancellation, $|\sum_k e^{2\pi i p_d(k)}| = \Theta(\sqrt N)$ (the typical situation), relative accuracy $\eta$ costs $O(\eta^{-1} \sqrt{N}\, \polylog N)$.
\end{corollary}

\begin{remark}[Which unweighted sums benefit]
\label{rem:weyl}
For $d \le 2$, classical $\polylog(N)$ algorithms already exist (Gauss sums admit closed forms, and the truncated theta function is handled by \cite{hiary2011nearly}), so Corollary~\ref{cor:weyl} offers no advantage. For $d \ge 3$, however, no classical algorithm essentially faster than direct $\Theta(N)$ summation is known for rigorous evaluation \cite{hiary2011fast}, and classical sampling (Remark~\ref{rem:montecarlo}) needs $\Theta(N/\eta^2)$ samples to resolve a $\Theta(\sqrt N)$-sized value to relative accuracy $\eta$---no better than direct summation. The quantum cost $O(\sqrt N)$ at fixed $\eta$ is then a quadratic improvement over the best known classical methods, for arbitrary degree. A further, cheaper regime: \emph{peak detection} in truncated Gauss sums, the primitive of Gauss-sum factoring proposals \cite{woelk2011factorization, merkel2011factorization}, asks only whether $N^{-1}|\sum_k e^{2\pi i f(k)}|$ is $\Theta(1)$ (candidate divisor) or $O(N^{-1/2})$ (non-divisor); this needs only constant additive precision, i.e., $O(1)$ amplitude-estimation iterations and $\polylog(N)$ gates per tested candidate. (This does not, of course, yield an efficient factoring algorithm, as the number of candidates is not reduced.)
\end{remark}

\section{Function-controlled rotations}
\label{sec:rot}

We now describe the implementation of the operator $U_g$ of \eqref{eq:uf} and its cost, in two variants.

\subsection{Via a fixed-point function register}
\label{subsec:register}
Under Assumption~\ref{ass:phase}, first compute the $b$-bit value
\be
\tilde g(k)=\sum_{i=-m_2}^{m_1} q_{k,i}\, 2^i , \qquad q_{k,i} \in \{0,1\},\quad b = m_1+m_2+1,
\label{eq:fk}
\ee
into an ancilla register $\ket{\tilde g(k)} = \ket{q_{k,m_1}, \cdots, q_{k,0} . q_{k,-1}, \cdots, q_{k,-m_2}}$. Since $\Rop(\alpha)\Rop(\beta)=\Rop(\alpha+\beta)$, the rotation by $\pi \tilde g(k)$ (more generally by $\pi \tilde g(k) \tau$ for any real constant $\tau$) factors into $b$ singly-controlled rotations, one per bit, with fixed angles $2^i \tau$: see Figure~\ref{fig:rotCircuit}. Uncomputing the register with a second oracle call completes $U_g$. The total cost is two phase-oracle calls plus $b$ controlled rotations. As shown in the proof of Theorem~\ref{thm:es}, $b = O(\log(1/\err))$ fractional bits suffice for final accuracy $\err$; only one integer bit is needed since the angle enters modulo $2$.

\begin{figure}
\centering
\[
\Qcircuit @C=1em @R=.7em {
\lstick{\ket{q_2}}    & \ctrl{5} & \qw      & \qw      & \qw      & \qw      & \qw \\
\lstick{\ket{q_1}}    & \qw      & \ctrl{4} & \qw      & \qw      & \qw      & \qw \\
\lstick{\ket{q_0}}    & \qw      & \qw      & \ctrl{3} & \qw      & \qw      & \qw \\
\lstick{\ket{q_{-1}}} & \qw      & \qw      & \qw      & \ctrl{2} & \qw      & \qw \\
\lstick{\ket{q_{-2}}} & \qw      & \qw      & \qw      & \qw      & \ctrl{1} & \qw \\
\lstick{\ket{0}}      & \gate{\Rop(2^2 \tau)} & \gate{\Rop(2^1 \tau)} & \gate{\Rop(\tau)} & \gate{\Rop(2^{-1} \tau)} & \gate{\Rop(2^{-2} \tau)} & \qw
}
\]
\caption{Rotation of an ancilla by angle $\pi g \tau$, where the fixed-point value $g$ is stored in a register as $\ket{q_2, q_1, q_0 . q_{-1}, q_{-2}}$ ($m_1 = m_2 = 2$) and $\tau$ is a real constant. Each register bit controls one rotation with a hard-wired angle.}
\label{fig:rotCircuit}
\end{figure}

\subsection{Directly, for polynomial phases}
\label{subsec:poly}
When $g$ is a polynomial $p_d(x)$ with real coefficients, the rotation can instead be applied directly by multi-controlled rotations, without a function register \cite{woerner2019quantum, stamatopoulos2019option}. Let $x=\sum_{i=-m_2}^{m_1} 2^i q_i$ be stored in the register $\ket{q_{m_1} \cdots q_{-m_2}}$ and consider $d=2$, $p_2(x) = a x^2 + b x + c$. Using $q_i^2 = q_i$ and $x^2 = \sum_i 2^{2i} q_i + \sum_{i<j} 2^{\,i+j+1} q_i q_j$,
\begin{align}
p_2(x) = c \;+\; \sum_{i} \bigl(2^{2i} a + 2^{i} b\bigr)\, q_i \;+\; \sum_{i<j} 2^{\,i+j+1} a \, q_i q_j .
\label{eq:polyexp}
\end{align}
Thus the rotation by $\pi p_2(x) \tau$ decomposes into one unconditional rotation ($c\tau$), one singly-controlled rotation per bit, and one doubly-controlled rotation per pair of bits; see Figure~\ref{fig:rot1Circuit}. For degree $d$ the analogous expansion produces $\sum_{r=0}^{d}\binom{m_1+m_2+1}{r} = O(b^d)$ multi-controlled rotations. For large $d$ or high precision the register method of Section~\ref{subsec:register}, whose cost is dominated by two evaluations of $p_d$ at $O(d\, b^2)$ gates each, is preferable; the direct method avoids the ancilla register and is convenient for small $d$.

\begin{figure}
\centering
\[
\Qcircuit @C=0.8em @R=.7em {
\lstick{\ket{q_2}}    & \ctrl{5} & \qw      & \qw      & \ctrl{5} & \qw      & \qw & \qw \\
\lstick{\ket{q_1}}    & \qw      & \ctrl{4} & \qw      & \qw      & \ctrl{4} & \qw & \qw \\
\lstick{\ket{q_0}}    & \qw      & \ctrl{3} & \ctrl{3} & \qw      & \qw      & \qw & \qw \\
\lstick{\ket{q_{-1}}} & \qw      & \qw      & \qw      & \ctrl{2} & \qw      & \qw & \qw \\
\lstick{\ket{q_{-2}}} & \qw      & \qw      & \qw      & \qw      & \ctrl{1} & \qw & \qw \\
\lstick{\ket{0}}      & \gate{\Rop((2^{4}a + 2^{2}b)\tau)} & \gate{\Rop(2^{2} a \tau)} & \gate{\Rop((a+b)\tau)} & \gate{\Rop(2^{2} a \tau)} & \gate{\Rop(a \tau)} & \gate{\Rop(c\tau)} & \qw
}
\]
\caption{Direct rotation by $\pi(a x^2+b x+ c)\tau$ for $x$ stored as $\ket{q_2, q_1, q_0 . q_{-1}, q_{-2}}$, following \eqref{eq:polyexp}. Shown are a representative subset: two of the five singly-controlled rotations (controls $q_2$ and $q_0$, angles $(2^{2i}a+2^i b)\tau$ for $i = 2, 0$), three of the ten doubly-controlled rotations (pairs $(q_1,q_0)$, $(q_2,q_{-1})$ and $(q_1,q_{-2})$, angles $2^{i+j+1}a\tau$), and the unconditional rotation by $c\tau$.}
\label{fig:rot1Circuit}
\end{figure}

\section{Classical formulas for \texorpdfstring{$\zeta(s)$}{zeta(s)}}
\label{sec:riemann}

We collect the two classical evaluation formulas whose main sums our quantum algorithm accelerates, together with the remainder bounds needed for a rigorous total error budget.

\subsection{Euler--Maclaurin summation}
For $g \in C^{2K}([a,b])$ and integers $a<b$ \cite[\S 2.10]{rubinstein2005computational},
\begin{align}
\sum_{n=a}^b g(n) = \int_a^b g(x)\, dx + \frac{g(a)+g(b)}{2}
+ \sum_{k=1}^{K} \frac{B_{2k}}{(2k)!}\Bigl( g^{(2k-1)}(b)- g^{(2k-1)}(a)\Bigr) + R_K,
\label{eq:em-general}
\end{align}
with
\ben
R_K = \int_a^b \frac{B_{2K} - B_{2K}(\{x\})}{(2K)!}\, g^{(2K)}(x)\, dx,
\qquad
|R_K| \le \frac{4\, \zeta(2K)}{(2\pi)^{2K}} \int_a^b \bigl| g^{(2K)}(x) \bigr|\, dx .
\een
Here $B_k(x)$ are the Bernoulli polynomials, defined by $z e^{z x}/(e^z-1) = \sum_{k \ge 0} B_k(x) z^k / k!$, $B_k = B_k(0)$ are the Bernoulli numbers (so $B_1 = -1/2$ and $B_{2k+1}=0$ for $k \ge 1$), $\{x\}$ is the fractional part, and the remainder bound uses $|B_{2K}(\{x\})| \le |B_{2K}| = 2 (2K)!\, \zeta(2K) / (2\pi)^{2K}$.

Applied to the tail of the zeta series, \eqref{eq:em-general} gives, for any integers $N \ge 1$ and even $K \ge 2$, and all $s \ne 1$ by analytic continuation \cite{rubinstein2005computational, edwards1974riemannaes},
\be
\zeta(s) = \sum_{n=1}^{N} n^{-s} + \frac{N^{1-s}}{s-1} - \frac{N^{-s}}{2}
+ \sum_{k=1}^{K/2} \frac{B_{2k}}{(2k)!} \Bigl( \prod_{j=0}^{2k-2}(s+j) \Bigr) N^{-s-2k+1}
+ E_K(s, N),
\label{eq:em}
\ee
where \cite{rubinstein2005computational}
\be
|E_K(s,N)| \;\le\; \frac{\zeta(K)}{\pi N^{\sigma}}\, \frac{|s+K-1|}{\sigma+K-1} \prod_{j=0}^{K-2} \frac{|s+j|}{2 \pi N}.
\label{eq:embound}
\ee
If $N$ is chosen so large that $|s+j|/(2\pi N) \le 1/10$ for $0 \le j \le K-2$, and $\sigma \ge 1/2$, then $|E_K| \le 10^{-(K-1)} |s+K-1|/(\sigma+K-1)$, so $D$ decimal digits are guaranteed once
\be
K-1 \;\ge\; D + \log_{10} \frac{|s+K-1|}{\sigma + K -1}.
\ee
The constraint $N = \Omega(|s|) = \Omega(t)$ is what limits Euler--Maclaurin at large height: the main sum then has $\Theta(t)$ terms. For $s$ near the real axis the formula converges extremely fast ($N \approx 2K$, $K \approx D+1$ suffice), and we use exactly this to evaluate the prefix sums $H_M(\sigma) = \sum_{k \le M} k^{-\sigma}$ (take $s = \sigma$ real in \eqref{eq:em} with the sum cut at $M$) to precision $10^{-D}$ in $\mathrm{poly}(D, \log M)$ time, as required by Assumption~\ref{ass:weights}.

\subsection{The Riemann--Siegel formula}
For $s$ in the critical strip the RS formula reads \cite{edwards1974riemannaes}
\begin{equation}
\zeta ( s ) = I_{0}( s ) + \chi ( s )\, \overline{I_{0}( 1-\overline{s}) },
\qquad
\chi(s)=\pi^{s-\frac{1}{2}}\, \frac{\Gamma\bigl(\frac{1-s}{2}\bigr)}{\Gamma\bigl(\frac{s}{2}\bigr)},
\label{eq:z2}
\end{equation}
where the bar denotes complex conjugation and, for integer $N \ge 0$,
\begin{align}
I_{N}( s ) = \int_{N\nearrow N+1} \frac{e^{i\pi z^{2}}\, z^{-s}}{e^{i\pi z} -e^{-i\pi z}}\, dz,
\label{eq:IN}
\end{align}
the contour being a straight line of direction $e^{i\pi/4}$ crossing the real axis between $N$ and $N+1$ (from the third quadrant to the first). The residue theorem gives $I_{N}(s) = I_{N-1}(s) - N^{-s}$, whence
\begin{equation}
I_{0}( s ) = \sum_{k=1}^{N} k^{-s} + I_{N}( s ) .
\label{eq:I0}
\end{equation}
Choosing $N=\floor{\sqrt{t/(2\pi)}}$ places a saddle point of the integrand on the contour, and the saddle-point expansion of $I_N$ in inverse powers of $t^{1/4}$ has rigorously bounded remainders: Gabcke \cite{gabcke1979neue} for the first several orders, and Arias de Reyna \cite{arias2011high} for all orders, with the error after $m$ correction terms bounded by $C_m\, t^{-(2m+3)/4}$ for explicit constants $C_m$. Consequently, for any fixed $\lambda > 0$, the non-sum part of \eqref{eq:z2}--\eqref{eq:I0} (that is, $\chi$, the correction terms, and the analogous quantities for $1-\bar s$) can be computed classically to within $t^{-\lambda}$ in $\polylog(t)$ time, with $O(\lambda)$ correction terms. What remains---and what dominates all classical algorithms---are the main sums $\sum_{k \le N} k^{-s}$ and $\sum_{k \le N} k^{-(1-\bar s)}$ with $N = \Theta(t^{1/2})$ terms.

\section{The Riemann zeta function on a quantum computer}
\label{sec:qc}

\subsection{The main sum as a weighted exponential sum}
Write the RS main sum as
\begin{align}
T_N(s) = \sum_{k=1}^{N} k^{-s} = \sum_{k=1}^{N} k^{-\sigma}\, e^{ -i t \ln k }
= H_N(\sigma) \sum_{k=1}^{N} w_k\, e^{2 \pi i f(k)},
\label{eq:mainsum}
\end{align}
with
\ben
w_k = \frac{k^{-\sigma}}{H_N(\sigma)}, \qquad
H_M(\sigma) = \sum_{k=1}^{M} k^{-\sigma}, \qquad
f(k) = -\frac{t \ln k}{2\pi} .
\een
The weights satisfy Assumption~\ref{ass:weights}: for $M$ held in a quantum register, $H_M(\sigma)$ is computable to $b$ bits by a reversible circuit implementing the Euler--Maclaurin formula \eqref{eq:em} truncated at $O(b)$ terms, each term a fixed-point power $M^{-\sigma - 2k+1}$ evaluated via $\exp$ and $\ln$ routines \cite{haner2018optimizing, wang2020quantum}; the cost is $\mathrm{poly}(b, \log t)$ gates per evaluation, and per Remark~\ref{rem:anatomy} this coherent arithmetic---not the amplitude estimation itself---dominates the $\polylog$ factor per Grover iterate. Hence the state $\sum_k \sqrt{w_k} \ket{k}$ is preparable by Lemma~\ref{lem:gr}; the normalization $H_N(\sigma)$ is needed only classically, to $O(\delta/H_N)$ relative precision, again by \eqref{eq:em}. The phase satisfies Assumption~\ref{ass:phase}: to know $f(k) \bmod 2$ to $b$ bits one needs $\ln k$ to absolute precision $2\pi\,2^{-b}/t$, i.e., $O(b + \log t)$ correct bits, computable in $\mathrm{poly}(b, \log t)$ gates \cite{wang2020quantum, haner2018optimizing}; the subsequent multiplication by $t/2\pi$ and reduction modulo $2$ are exact fixed-point operations on $O(b+\log t)$-bit registers, and no trigonometric evaluation is required (Section~\ref{subsec:register}). Theorem~\ref{thm:es} and Corollary~\ref{cor:unnormalized} therefore estimate $T_N(s)$ to additive error $\delta$ at cost $O\bigl( (H_N(\sigma)/\delta)\, \polylog(t/\delta)\log(1/\gamma) \bigr)$, and for $0<\sigma<1$,
\be
H_N(\sigma) \le 1 + \frac{N^{1-\sigma}}{1-\sigma} = O_\sigma\bigl( t^{(1-\sigma)/2} \bigr), \qquad N = \floor{\sqrt{t/2\pi}} .
\label{eq:HN}
\ee

\subsection{Main theorem}

\begin{theorem}
\label{thm:zeta}
Fix $\sigma \in (0,1)$ and $\lambda > 0$. There is a quantum algorithm which, given $t \ge 2$, $\gamma \in (0,1)$ and $\delta \in [t^{-\lambda}, 1)$, outputs $\widehat\zeta$ with
\ben
\bigl| \widehat\zeta - \zeta(\sigma + i t) \bigr| \le \delta
\quad\text{with probability at least } 1-\gamma,
\een
at a total gate cost
\ben
O_{\sigma,\lambda}\!\left( \frac{t^{(1-\sigma)/2}}{\delta}\; \polylog(t/\delta)\, \log(1/\gamma) \right).
\een
In particular, $\zeta(1/2+it)$ is estimated to accuracy $\delta$ in $O(t^{1/4}\, \delta^{-1} \polylog(t/\delta) \log(1/\gamma))$ gates.
\end{theorem}

\begin{proof}
Use \eqref{eq:z2} and \eqref{eq:I0} with $N=\floor{\sqrt{t/(2\pi)}}$:
\ben
\zeta(s) = T_N(s) + \chi(s) \overline{T_N(1 - \bar s)} + \Delta(s),
\een
where $\Delta$ collects $I_N(s)$ and $\chi(s)\overline{I_N(1-\bar s)}$. By Section~\ref{sec:riemann}, $\Delta(s)$ and $\chi(s)$ are computable classically to within $\delta/4$ in $\polylog(t/\delta)$ time, using $O(\lambda)$ RS correction terms with the rigorous bounds of \cite{gabcke1979neue, arias2011high} (and Stirling-series evaluation of $\chi$).

The two main sums are handled by Corollary~\ref{cor:unnormalized}. For $T_N(s)$: the $\ell^1$ mass is $H_N(\sigma) = O(t^{(1-\sigma)/2})$ by \eqref{eq:HN}, and we demand additive error $\delta/4$, at cost $O((H_N(\sigma)/\delta)\polylog(t/\delta)\log(1/\gamma)) = O(t^{(1-\sigma)/2}\delta^{-1}\polylog(t/\delta)\log(1/\gamma))$. For the reflected sum $T_N(1-\bar s) = \sum_{k \le N} k^{-(1-\sigma)} e^{-it\ln k}$: its $\ell^1$ mass is $H_N(1-\sigma) = O(t^{\sigma/2})$, and since it enters multiplied by $\chi(s)$, with $|\chi(\sigma+it)| = (t/2\pi)^{1/2-\sigma}(1+O(1/t))$ \cite{titchmarsh1986theory}, it must be estimated to additive error $\delta / (4|\chi(s)|)$; the cost is
\ben
O\Bigl( \frac{H_N(1-\sigma)\, |\chi(s)|}{\delta} \polylog \Bigr)
= O\Bigl( \frac{t^{\sigma/2}\; t^{1/2-\sigma}}{\delta} \polylog \Bigr)
= O\Bigl( \frac{t^{(1-\sigma)/2}}{\delta} \polylog \Bigr),
\een
the same as the first sum. On the critical line the two sums coincide, since $1-\bar s = s$ there. Summing the four error contributions ($2 \times \delta/4$ from the quantum estimates weighted as above, $\delta/4$ from $\Delta$, $\delta/4$ from classical evaluation of $\chi$ and roundoff) gives total error $\le \delta$, and a union bound over the amplitude estimations gives failure probability $\le \gamma$ after the median amplification of Lemma~\ref{lem:median}.
\end{proof}

\begin{remark}[Where the quantum advantage lies, and its limits]
\label{rem:crossover}
At $\sigma = 1/2$ and accuracy $\delta = t^{-\lambda}$ the cost of Theorem~\ref{thm:zeta} is $t^{1/4+\lambda+o(1)}$. Compared with classical algorithms:
the RS formula costs $t^{1/2+o(1)}$, so the quantum algorithm achieves a strict improvement whenever $\lambda < 1/4$;
Hiary's algorithm \cite{hiary2011fast} costs $t^{4/13+o(1)}$ with only polynomial dependence on $\lambda$, so the quantum algorithm is advantageous precisely when $\lambda < 4/13 - 1/4 = 3/52$. Thus for very high precision ($\delta \ll t^{-3/52}$) classical methods remain superior, a direct consequence of the $\delta^{-1}$ (Heisenberg-limited) precision scaling of amplitude estimation; and for fixed or mildly shrinking accuracy---the regime relevant to locating zeros---the quantum algorithm gives the best known scaling, $t^{1/4+o(1)}$. This also corrects a tempting overstatement: since the normalized sum in \eqref{eq:mainsum} is estimated to additive precision, a $\polylog(t)$-time quantum evaluation of $\zeta$ to fixed accuracy does \emph{not} follow, because of the $\ell^1$-mass factor $H_N(\sigma) = \Theta(t^{(1-\sigma)/2})$ (Remark~\ref{rem:l1}).
\end{remark}

\begin{remark}[Verifying RH: one amplitude estimation per point]
\label{rem:zeros}
For locating zeros on the critical line one works with the real Riemann--Siegel $Z$-function $Z(t) = e^{i\theta(t)} \zeta(1/2+it)$, where $\theta(t) = \arg \Gamma(1/4 + it/2) - \tfrac{t}{2}\ln \pi$ is computable classically via the Stirling series. The RS formula gives
\ben
Z(t) = 2 \sum_{k=1}^{N} k^{-1/2} \cos\bigl( \theta(t) - t \ln k \bigr) + R(t), \qquad N = \floor{\sqrt{t/2\pi}},
\een
with $R(t)$ classically computable to $t^{-\Theta(m)}$ from $m$ correction terms \cite{gabcke1979neue, arias2011high}. The sum is $2 H_N(1/2) \,\bigl(\sum_k w_k \cos 2\pi f(k)\bigr)$ with $f(k) = (\theta(t) - t\ln k)/2\pi$, so by \eqref{eq:recover} a \emph{single} amplitude estimation per evaluation point suffices, at cost $O(t^{1/4}\delta^{-1}\polylog t)$. Sign changes of $Z$ certify zeros on the critical line, and Turing's method bounds the zero count using evaluations of moderate accuracy $\delta \sim 1/\log t$ \cite{edwards1974riemannaes, titchmarsh1986theory}, so the per-zero verification cost is $t^{1/4+o(1)}$, against $t^{4/13+o(1)}$ classically \cite{hiary2011fast, bober2018new}.
\end{remark}

\begin{remark}[Euler--Maclaurin variant]
The same construction applied to \eqref{eq:em} with $N = \Theta(t)$ estimates $\zeta(\sigma+it)$ at cost $O(t^{1-\sigma} \delta^{-1} \polylog(t/\delta))$, worse than Theorem~\ref{thm:zeta} but with completely elementary remainder bounds \eqref{eq:embound}; it may serve as an independent check in implementations.
\end{remark}

\subsection{Block decompositions do not help the quantum algorithm}
Hiary's classical speedups \cite{hiary2011fast} partition the main sum into $R = t^{1/2-\beta}$ consecutive blocks of length $t^{\beta}$, approximate each block by exponential sums with polynomial phase of degree $d = \floor{1/(1/2 - \beta)} - 1$ (Taylor-expanding $t \ln k$ about the block midpoint), and evaluate blocks by fast theta-type algorithms. Since Theorem~\ref{thm:es} evaluates polynomial-phase sums as easily as logarithmic-phase ones, one might hope to combine blocking with quantum estimation and push the exponent below $1/4$. This fails:

\begin{proposition}
\label{prop:blocks}
Partition $\{1, \dots, N\}$ into blocks $j = 1, \dots, R$, and let $h_j > 0$ be the $\ell^1$ mass $\sum_{k \in \mathrm{block}\ j} k^{-\sigma}$, so $\sum_j h_j = H_N(\sigma)$. Any strategy that estimates each block sum by amplitude estimation to additive error $h_j \err_j$ and adds the results, subject to worst-case total error $\sum_j h_j \err_j \le \delta$, uses at least
\ben
\sum_{j=1}^R \Omega(1/\err_j) \;=\; \Omega\!\left( \frac{\bigl( \sum_j \sqrt{h_j} \bigr)^2}{\delta} \right) \;\ge\; \Omega\!\left( \frac{H_N(\sigma)}{\delta} \right)
\een
Grover iterations, with equality on the right iff $R = 1$. Hence blocking can only increase the worst-case query complexity, by up to a factor $R$.
\end{proposition}

\begin{proof}
Amplitude estimation to additive error $\err_j$ on a normalized amplitude requires $\Omega(1/\err_j)$ iterations \cite{brassard2002quantum}. Minimizing $\sum_j 1/\err_j$ subject to $\sum_j h_j \err_j = \delta$ (Lagrange multipliers, or Cauchy--Schwarz in the form $\sum 1/\err_j \cdot \sum h_j \err_j \ge (\sum \sqrt{h_j})^2$) gives the optimum $(\sum_j \sqrt{h_j})^2/\delta$ at $\err_j \propto h_j^{-1/2}$. Finally $(\sum_j \sqrt{h_j})^2 \ge \sum_j h_j = H_N(\sigma)$, with equality iff a single $h_j$ is nonzero.
\end{proof}

Blocking may still be useful at the \emph{gate} level---polynomial phases avoid the fixed-point logarithm---but it cannot improve the $t^{1/4}$ query scaling, which is already the square root of the $t^{1/2}$ term count that blocking was invented to beat classically. Note also that Proposition~\ref{prop:blocks} makes the implementation cost of the block oracles irrelevant to this conclusion: the lower bound holds even if each block's state preparation and polynomial-phase oracle were granted at unit cost.

\section{Magnitude of a sum of quantum amplitudes}
\label{sec:mag}

We record a general tool of independent interest: estimating the magnitude of the sum of the amplitudes of an arbitrary efficiently preparable state. For exponential sums it estimates the \emph{amplitude-weighted} sum $\sum_k \sqrt{w_k}\, e^{2\pi i f(k)}$ (contrast the probability-weighted sum of Theorem~\ref{thm:es}), though only in magnitude and at a normalization penalty that we make explicit.

\begin{proposition}
\label{prop:mag}
Let $P$ be a circuit preparing $\ket{\psi}_n = \sum_{k=0}^{N-1} a_k \ket{k}_n$, $\sum_k |a_k|^2 = 1$, and set
\ben
\mu \;=\; \Bigl| \braket{u | \psi} \Bigr| \;=\; \frac{1}{\sqrt N}\, \Bigl| \sum_{k=0}^{N-1} a_k \Bigr| \;\in [0,1],
\qquad \ket{u} = \frac{1}{\sqrt N}\sum_k \ket{k}_n .
\een
Then $\mu$ can be estimated to additive error $\err$ with probability $1-\gamma$ using $O(\err^{-1}\log(1/\gamma))$ applications of $P$ and $P^{-1}$ and $O(n)$ additional gates per application.
\end{proposition}

\begin{proof}
Let $\mathcal{A} = F \cdot (H^{\otimes n} \otimes I) \cdot (P \otimes I)$ act on $\ket{0}_n \ket{0}$, where the flag $F$ maps $\ket{x}_n\ket{0} \mapsto \ket{x}_n\ket{x \ne 0^n}$ (an $(n{-}1)$-fold controlled NOT on zero-controls, cost $O(n)$). Since the amplitude of $\ket{0^n}$ in $H^{\otimes n}\ket{\psi}$ is $\braket{u|\psi} = N^{-1/2}\sum_k a_k$,
\ben
\mathcal{A} \ket{0}_n \ket{0} = \braket{u | \psi}\, \ket{0^n}\ket{0} + \sqrt{1 - \mu^2}\; \ket{\psi^\perp}\ket{1}
\een
for a normalized $\ket{\psi^\perp}$. The good probability is $\mu^2$; amplitude estimation (Theorem~\ref{thm:ae}) with $M = O(1/\err)$ estimates $p=\mu^2$ to $O(\err \sqrt{p(1-p)} + \err^2)$, hence $\mu = \sqrt{p}$ to $O(\err)$ (the map $p \mapsto \sqrt p$ contracts absolute error for $p \ge \err^2$; for $p < \err^2$ both $\tilde\mu, \mu \le 2\err$). Median amplification gives the confidence claim.
\end{proof}

\begin{remark}[Equivalent form: inversion about the mean]
The reflection $D = H^{\otimes n} (2\ket{0^n}\bra{0^n} - I) H^{\otimes n}$ implements Grover's inversion about the mean \cite{grover1996fast, nielsen2002quantum}: $D$ maps amplitudes $a_k \mapsto 2\bar a - a_k$ with $\bar a = N^{-1}\sum_k a_k$. Appending an ancilla and flagging as above, one finds the state
\ben
\sqrt{N}\, |\bar a| \, \ket{u}\ket{0} \;+\; \sqrt{1 - N |\bar a|^2}\, \ket{\psi_1}\ket{1},
\qquad \sqrt N |\bar a| = \mu,
\een
recovering Proposition~\ref{prop:mag}; the two derivations are related by $D$'s factorization through $H^{\otimes n}$.
\end{remark}

\begin{remark}[Application to exponential sums, and cost accounting]
\label{rem:magcost}
Take $P = U_{\mathrm{ph}} W$, where $W$ prepares $\sum_k \sqrt{w_k}\ket{k}$ (Lemma~\ref{lem:gr}) and $U_{\mathrm{ph}} : \ket{k} \mapsto e^{2\pi i f(k)}\ket{k}$ is a diagonal phase, implemented with the circuits of Section~\ref{sec:rot} with $\Rop$ replaced by phase rotations. Then $a_k = \sqrt{w_k}\, e^{2\pi i f(k)}$ and Proposition~\ref{prop:mag} estimates
\ben
\mu = \frac{1}{\sqrt N} \Bigl| \sum_{k=0}^{N-1} \sqrt{w_k}\, e^{2\pi i f(k)} \Bigr| .
\een
Consequently $\bigl|\sum_k \sqrt{w_k}e^{2\pi i f(k)}\bigr|$ is obtained to additive error $\delta$ at cost $O(\sqrt N / \delta)$---the factor $\sqrt N$ is again the denormalization price of Remark~\ref{rem:l1} and is unavoidable in general. For uniform weights $w_k = 1/N$ the two weightings coincide ($\mu = |S(f, \mathrm{unif}, N)|$) and the cost matches Theorem~\ref{thm:es}, which additionally provides the phase of the sum.
\end{remark}

\begin{remark}[Magnitude route to the zeta main sum: an independent consistency check]
\label{rem:magzeta}
Proposition~\ref{prop:mag} gives a second, structurally different path to the Riemann--Siegel main sum \eqref{eq:mainsum}. Take amplitudes proportional to the \emph{full} coefficient, $a_k = k^{-\sigma} e^{-it\ln k} / \sqrt{H_N(2\sigma)}$ for $1 \le k \le N$ (preparable by Lemma~\ref{lem:gr} with weights $\propto k^{-2\sigma}$, followed by the diagonal phase as in Remark~\ref{rem:magcost}). Then
\ben
\mu = \frac{\bigl| T_N(s) \bigr|}{\sqrt{N\, H_N(2\sigma)}},
\een
and Proposition~\ref{prop:mag} yields $|T_N(s)|$ to additive error $\delta$ at cost $O\bigl(\sqrt{N H_N(2\sigma)}/\delta\bigr)$. By Cauchy--Schwarz, $H_N(\sigma) \le \sqrt{N H_N(2\sigma)}$, so this route is never better than that of Theorem~\ref{thm:zeta}; at $\sigma = 1/2$ it is worse only by a logarithmic factor, since $\sqrt{N H_N(1)} = O(t^{1/4}\sqrt{\log t})$ against $H_N(1/2) = O(t^{1/4})$. That two different encodings---an $\ell^1$-based one (probability weights, Theorem~\ref{thm:es}) and an $\ell^2$-based one (amplitude weights, Proposition~\ref{prop:mag})---meet at the same exponent $1/4$ is a useful consistency check on Theorem~\ref{thm:zeta}. The magnitude route returns $|T_N|$ only, with no phase information, so it cannot drive the sign-change certification of zeros in Remark~\ref{rem:zeros}.
\end{remark}

\section{Amplitude and phase estimation}
\label{sec:phase}

For completeness we review the estimation subroutines used above; see \cite{brassard2002quantum, nielsen2002quantum, kitaev2002classical, svore2013faster, wiebe2016efficient, mohammadbagherpoor2019improved, suzuki2020amplitude, aaronson2020quantum} for details and variants.

\subsection{Amplitude estimation as phase estimation}
Let $\mathcal{A}$ be as in \eqref{eq:ampl}, with good probability $p = \sin^2 \theta_p$. Define the two reflections
\ben
S_{0} = I - 2 \ket{0}_{n+1}\bra{0}_{n+1},
\qquad
S_{\chi} = I - 2\, \Pi_{\mathrm{good}},
\een
where $\Pi_{\mathrm{good}}$ projects onto the flagged (good) subspace. The Grover iterate
\be
Q = - \mathcal{A}\, S_0\, \mathcal{A}^{-1} S_{\chi}
\label{eq:q}
\ee
acts in the two-dimensional space spanned by $\ket{\Psi_0}, \ket{\Psi_1}$ as a rotation by $2\theta_p$ \cite{brassard2002quantum}:
\be
Q^m \mathcal{A}\ket{0}_{n+1} = \sin\bigl((2m+1) \theta_p\bigr) \ket{\Psi_1} + \cos\bigl((2m+1)\theta_p\bigr) \ket{\Psi_0}.
\label{eq:qm}
\ee
The eigenvalues of $Q$ on this subspace are $e^{\pm 2 i \theta_p}$, so estimating $p$ reduces to estimating the phase of a unitary---the problem addressed below---and Theorem~\ref{thm:ae} follows by applying QFT-based phase estimation to $Q$.

\subsection{QFT-based phase estimation}
\label{subsec:qft}
Given a unitary $U$ with eigenpair $U\ket{\xi} = e^{2\pi i \varphi}\ket{\xi}$, $\varphi \in [0,1)$, adjoin an $n$-qubit index register prepared by Hadamards in $N^{-1/2}\sum_{l<N} \ket{l}$ and apply the controlled power $\Lambda(U): \ket{l}\ket{\xi} \mapsto \ket{l} U^{l} \ket{\xi}$. Writing $l = \sum_{i=0}^{n-1} l_i 2^i$ in binary,
\begin{align}
U^{l} = \prod_{i=0}^{n-1} \bigl( U^{2^{i}} \bigr)^{l_i},
\end{align}
so $\Lambda(U)$ is a product of $n$ singly-controlled powers $U^{2^i}$, as in Figure~\ref{fig:phase1}. The register is left in $N^{-1/2}\sum_l e^{2\pi i \varphi l}\ket{l}$, and the inverse quantum Fourier transform followed by measurement returns the best $n$-bit approximation of $\varphi$ with probability at least $8/\pi^2$ \cite{nielsen2002quantum}. Precision $2^{-n}$ requires $2^n - 1$ applications of $U$: accuracy $\err$ costs $\Theta(1/\err)$ applications, the Heisenberg limit. Exponential accuracy in the qubit count is therefore available only when high powers $U^{2^i}$ admit shortcuts, as modular exponentiation does in Shor's algorithm \cite{shor1999polynomial}; for the iterates $Q$ of \eqref{eq:q} no such shortcut is known, which is why the $\delta^{-1}$ factors in this paper cannot be reduced to $\polylog(1/\delta)$ by known techniques.

\begin{figure}[htb]
  \centering
  \[
 \Qcircuit @C=.7em @R=.7em {
  \lstick{\ket{0}}    & \gate{H} & \qw & \ctrl{4} & \qw        & \qw & \cdots & & \qw                & \multigate{3}{QFT^{-1}} & \qw    & \meter & \cw \\
  \lstick{\ket{0}}    & \gate{H} & \qw & \qw      & \ctrl{3}   & \qw & \cdots & & \qw                & \ghost{QFT^{-1}}	& \qw  & \meter & \cw \\
  \lstick{\vdots\ \ } & \vdots   &     &          &            &     &        & &                    & \pureghost{QFT^{-1}}    &        & \vdots &     \\
  \lstick{\ket{0}}    & \gate{H} & \qw & \qw      & \qw        & \qw & \cdots & & \ctrl{1}           & \ghost{QFT^{-1}}	& \qw  & \meter & \cw \\
  \lstick{\ket{\xi}} & /^n \qw  & \qw & \gate{U^{2^{n-1}}} & \gate{U^{2^{n-2}}}  & \qw & \cdots & & \gate{U^{2^{0}}} & \qw
 }
  \]
  \caption{QFT-based phase (eigenvalue) estimation.}
  \label{fig:phase1}
\end{figure}

\subsection{Estimation with classical post-processing}
\label{subsec:classical}
The QFT can be dispensed with entirely \cite{kitaev1995quantum, svore2013faster, wiebe2016efficient, suzuki2020amplitude, aaronson2020quantum}. The basic primitive uses one control qubit, a phase gate
$Z(\theta) = \mathrm{diag}(1, e^{i\theta})$,
and a controlled power $U^M$, as in Figure~\ref{fig:measCircuit}. A direct computation shows the control qubit ends in
\be
\frac{1+e^{i(2\pi M \varphi+\theta)}}{2}\,\ket{0} + \frac{1-e^{ i (2\pi M \varphi+\theta)}}{2}\,\ket{1},
\ee
so the outcome probabilities are
\be
\label{eq:cmp}
P_{M,\theta}(0) = \frac{1 + \cos(2\pi M \varphi + \theta)}{2},
\qquad
P_{M,\theta}(1) = \frac{1 - \cos(2\pi M \varphi + \theta)}{2}.
\ee

\begin{figure}
\centering
\[
\Qcircuit @C=1em @R=1em {
\lstick{\ket{0}} & \gate{H}  & \gate{Z(\theta)}   &\ctrl{1}   & \gate{H}    & \meter    & \cw \\
\lstick{\ket{\xi}} &   {/^n} \qw   &\qw   & \gate{U^M} &    \qw      &    \qw    & \qw }
\]
\caption{Measurement primitive for QFT-free phase estimation.}
\label{fig:measCircuit}
\end{figure}

Repeated runs at $M=1$, $\theta = 0$ estimate $\cos(2\pi\varphi)$, which determines $\varphi$ only up to the reflection $\varphi \leftrightarrow 1 - \varphi$; runs at $\theta = \pi/2$ estimate $\sin(2\pi\varphi)$ and resolve the ambiguity. With $O(\err^{-2}\log(1/\gamma))$ samples this gives $\varphi$ to accuracy $\err$; using increasing powers $M = 1, 2, 4, \dots$ as in Kitaev's algorithm \cite{kitaev1995quantum}---each doubling sheds one bit of the unknown integer multiple of $2\pi$ and pins down one further bit of $\varphi$---or the maximum-likelihood and iterative schemes of \cite{svore2013faster, wiebe2016efficient, suzuki2020amplitude}, the total number of applications of $U$ to reach accuracy $\err$ with confidence $1 - \gamma$ is again $O(\err^{-1}\log(1/\gamma))$. In the same setting Aaronson and Rall \cite{aaronson2020quantum} give a QFT-free amplitude estimator with multiplicative guarantee: an estimate $\hat p$ with $(1-\err) p \le \hat p \le (1+\err) p$ using $O\bigl(\err^{-1} p^{-1/2}\bigr)$ applications of $\mathcal{A}^{\pm 1}$, with the usual $\log(1/\gamma)$ overhead for confidence $1-\gamma$. These QFT-free variants involve only one control qubit and shallow circuits between measurements, and are the natural choice for early fault-tolerant hardware; all statements in Sections~\ref{sec:gen}--\ref{sec:mag} hold verbatim with any of them in place of Theorem~\ref{thm:ae}.

\section{Discussion and conclusion}
\label{sec:conclusion}

We have given a rigorous account of the quantum evaluation of weighted exponential sums by amplitude estimation, under explicit oracle assumptions (efficient prefix sums for the weights; fixed-point circuits for the phase), and applied it to the Riemann zeta function. The resulting algorithm estimates $\zeta(\sigma+it)$ to accuracy $\delta$ in $O(t^{(1-\sigma)/2}\delta^{-1}\polylog(t/\delta))$ gates---$t^{1/4}$ on the critical line---which, in the fixed-accuracy regime relevant to locating and counting zeros, improves on the best rigorous classical algorithm's $t^{4/13+o(1)}$ \cite{hiary2011fast}. The improvement is exactly a quadratic (Grover-type) speedup over the Riemann--Siegel main-sum length, and we have shown two structural limits: the $\ell^{1}$ mass of the main sum is an unavoidable multiplier of the amplitude-estimation error, which rules out a $\polylog(t)$ algorithm by these techniques, and block decompositions of the main sum cannot improve the quantum query complexity (Proposition~\ref{prop:blocks}). Because amplitude estimation is Heisenberg-limited, the $\delta^{-1}$ accuracy scaling means classical methods remain superior when extreme precision $\delta \ll t^{-3/52}$ is required; for zero verification, where accuracy $\delta \sim 1/\log t$ suffices for sign changes of $Z(t)$ and for Turing's method, the quantum scaling $t^{1/4+o(1)}$ per point stands (Remark~\ref{rem:zeros}).

The estimates here are asymptotic and oracle-based; a resource count at the level of Toffoli gates and logical qubits for a concrete height $t$---including the fixed-point logarithm, the Grover--Rudolph angles from Euler--Maclaurin prefix sums, and the RS correction terms---is a natural next step, as is the question of whether structured phases (e.g., quadratic, where classical $\polylog$ algorithms exist \cite{hiary2011nearly}) admit quantum evaluation with better than $\delta^{-1}$ precision scaling. Extensions of the present analysis to Dirichlet $L$-functions and other Dirichlet series with efficiently computable coefficient prefix sums are straightforward.


\bigskip
\bibliographystyle{abbrv}
\bibliography{quantum}

\end{document}